\documentclass[12pt,a4paper]{article}

\advance \textheight by \topskip
\usepackage{fullpage}
\usepackage[centertags]{amsmath}
\usepackage{amsbsy}
\usepackage{amsfonts}
\usepackage{amssymb}
\usepackage[dvips]{graphicx}
\usepackage[dvips,monochrome]{color}
\allowdisplaybreaks[1]

\newcommand{\openone}{\leavevmode\hbox{\small1\kern-3.8pt\normalsize1}}

\newcommand{\vet}[1]{\ensuremath{\hskip-1pt\vec{\hskip1pt#1}}}

\begin{document}
\begin{flushright}
\begin{tabular}{l}
DFTT 15/2001
\\
hep-ph/0105319
\end{tabular}
\end{flushright}
\begin{center}
\large\bfseries
Quantum Mechanics of Neutrino Oscillations
\\[0.5cm]
\normalsize\normalfont
Carlo Giunti
\\
\small\itshape
INFN, Sezione di Torino,
\\
\small\itshape
and
\\
\small\itshape
Dipartimento di Fisica Teorica,
Universit\`a di Torino,
\\
\small\itshape
Via P. Giuria 1, I--10125 Torino, Italy
\end{center}
\begin{abstract}
Subtle problems in the theory of neutrino oscillations in vacuum
are discussed \cite{www-qmno}.
It is shown that Lorentz invariance implies that in general
flavor neutrinos in oscillation experiments
are superpositions of massive neutrinos with different energies
and different momenta.
It is argued that
a wave packet description of massive neutrinos
is necessary in order to understand the physics of neutrino oscillations.

Talk presented at the
XI$^{\mathrm{th}}$ International School ``PARTICLES and COSMOLOGY'',
18--24 April 2001,
Baksan Valley, Kabardino-Balkaria, Russian Federation.
\end{abstract}

\section{Introduction}
\label{Introduction}

The standard theory of neutrino oscillations in vacuum
\cite{Bilenky-Pontecorvo-PR-78}
has been developed in the 70's
on the basis of four main assumptions:
\renewcommand{\labelenumi}{\theenumi}
\renewcommand{\theenumi}{(A\arabic{enumi})}
\begin{enumerate}
\item \label{A1}
Neutrinos are extremely relativistic particles.
\item \label{A2}
Neutrinos produced in charged-current weak interaction processes
together with charged leptons $\alpha^+$ are described by the
flavor state
\begin{equation}
|\nu_{\alpha}\rangle
=
\sum_{k}
U_{{\alpha}k}^{*}
|\nu_{k}\rangle
\qquad
(\alpha=e,\mu,\tau)
\,,
\label{001}
\end{equation}
where $U$ is the mixing matrix of the neutrino fields,
$
\nu_{{\alpha}L}
=
\sum_{k}
U_{{\alpha}k}
\nu_{kL}
$.
The massive neutrino states
$ |\nu_{k}\rangle $
are quanta of the fields
$ \nu_{k} $
of neutrinos with mass $m_k$.
Their energy $E_k$ and momentum $p_k$ are connected by the relativistic
dispersion relation
\begin{equation}
E_k^2 = p_k^2 + m_k^2
\,.
\label{0011}
\end{equation}
Since in oscillation experiments neutrinos
propagate along a macroscopic distance
between production and detection,
we consider only one spatial direction
along the neutrino path.
\item \label{A3}
The massive neutrino states
$ |\nu_{k}\rangle $
have the same momentum $p$
(``equal momentum assumption'', $p_k=p$),
but different energies,
\begin{equation}
E_k
=
\sqrt{p^2 + m_k^2}
\simeq
p
+
\frac{m_k^2}{2p}
=
E
+
\frac{m_k^2}{2E}
\,.
\label{002}
\end{equation}
Here
$E=p$
is the energy of a massless neutrino
and the approximation is valid for extremely relativistic neutrinos.
\item \label{A4}
The time $T$ of propagation of neutrinos from source to detector
is approximately equal to the source-detector distance $L$.
\end{enumerate}
The massive neutrino states
$ |\nu_{k}\rangle $
evolve in time according to the Schr\"odinger equation,
whose solution is
\begin{equation}
|\nu_{k}(t)\rangle
=
e^{-iE_kt}
\,
|\nu_{k}\rangle
\,.
\label{003}
\end{equation}
Using the assumptions
\ref{A1}--\ref{A4}
it is straightforward to derive the
oscillation probability
\begin{equation}
P_{\nu_{\alpha}\to\nu_{\beta}}(T)
=
\left|
\sum_{k}
U_{{\alpha}k}^{*}
e^{-iE_kT}
U_{{\beta}k}
\right|^2
\label{004}
\end{equation}
as a function of the time $T$
and the oscillation probability
\begin{equation}
P_{\nu_{\alpha}\to\nu_{\beta}}(L)
=
\sum_{k}
|U_{{\alpha}k}|^2
|U_{{\beta}k}|^2
+
2
\mathrm{Re}
\sum_{k>j}
U_{{\alpha}k}^{*}
U_{{\beta}k}
U_{{\alpha}j}
U_{{\beta}j}^{*}
\exp\!\left(
-i\frac{\Delta{m}^2_{kj} L}{2E}
\right)
\label{005}
\end{equation}
as a function of the distance $L$,
with
$\Delta{m}^2_{kj} \equiv m_k^2 - m_j^2$.
The probability (\ref{005})
is the one observed in real experiments
where the propagation time $T$ in not measured
but the source-detector distance $L$ is known.

Let us examine critically the assumptions
\ref{A1}--\ref{A4}.

The assumption \ref{A1}
is easily understood to be correct
by noticing that,
although not all existing neutrinos are
extremely relativistic,
in neutrino oscillation experiments
one is interested in detectable neutrinos,
which have
energy larger than some fraction of MeV.
Indeed,
neutrinos are detected in:
\renewcommand{\labelenumi}{\theenumi.}
\renewcommand{\theenumi}{\arabic{enumi}}
\begin{enumerate}
\item
Charged current weak processes
which have an energy threshold
larger than some fraction of MeV.
For example\footnote{
In a scattering process
$\nu + A \to B + C$
the squared center-of-mass energy
$s = 2 E m_A + m_A^2$
must be bigger than
$( m_B + m_C )^2$,
leading to
$ \displaystyle
E_{\mathrm{th}}
=
\frac{ ( m_B + m_C )^2 }{ 2 m_A } - \frac{ m_A }{ 2 }
$.
}:
\begin{itemize}
\item
$
E_{\mathrm{th}}
=
0.233 \, \mathrm{MeV}
$
for
$ \nu_e + {}^{71}\mathrm{Ga} \to {}^{71}\mathrm{Ge} + e^- $
in the GALLEX \cite{GALLEX-99},
SAGE \cite{SAGE-99}
and
GNO \cite{GNO-00}
solar neutrino experiments.
\item
$
E_{\mathrm{th}}
=
0.81 \, \mathrm{MeV}
$
for
$ \nu_e + {}^{37}\mathrm{Cl} \to {}^{37}\mathrm{Ar} + e^- $
in the Homestake \cite{Homestake-98}
solar neutrino experiment.
\item
$
E_{\mathrm{th}}
=
1.8 \, \mathrm{MeV}
$
for
$ \bar\nu_e + p \to n + e^+ $
in reactor neutrino experiments
(for example Bugey \cite{Bugey} and CHOOZ \cite{CHOOZ-99}).
\item
$
E_{\mathrm{th}}
=
110 \, \mathrm{MeV}
$
for
$ \nu_\mu + n \to p + \mu^- $.
\item
$
E_{\mathrm{th}}
\simeq
m_\mu^2 / 2 m_e
=
10.9 \, \mathrm{GeV}
$
for
$ \nu_\mu + e^- \to \nu_e + \mu^- $.
\end{itemize}
\item
The elastic scattering process
$ \nu + e^- \to \nu + e^- $,
whose cross section is proportional to the neutrino energy
($
\sigma(E)
\sim
\sigma_0 E / m_e
$,
with
$
\sigma_0
\sim
10^{-44} \, \mathrm{cm}^2
$).
Therefore,
an energy threshold
of some MeV's
is needed in order to have a signal above the background.
For example,
$
E_{\mathrm{th}}
\simeq
5 \, \mathrm{MeV}
$
in the Super-Kamiokande \cite{SK-sun-01}
solar neutrino experiment.
\end{enumerate}
On the other hand,
although the direct experimental upper limits
for the effective neutrino masses
in lepton decays are not very stringent
($m_{\nu_e} \lesssim 3 \, \mathrm{eV}$,
$m_{\nu_\mu} \lesssim 190 \, \mathrm{keV}$,
$m_{\nu_\tau} \lesssim 18.2 \, \mathrm{MeV}$,
see Ref.~\cite{PDG}),
we know that
the sum of the masses of light neutrinos
($ m_\nu \lesssim m_Z/2 \simeq 45 \, \mathrm{GeV}$)
that have a substantial mixing with $\nu_e$, $\nu_\mu$ and $\nu_\tau$
is constrained to be smaller than about
$5 \, \mathrm{eV}$
by their contribution to the total energy density of the Universe
\cite{Tegmark-01}.

The comparison
of the cosmological limit on neutrino masses
with the energy threshold in the processes of neutrino detection
implies that detectable neutrinos are extremely relativistic.
This is an important fact that is crucial for the
theory of neutrino oscillations.

Let us consider now the assumption \ref{A2}.
The flavor state
$ |\nu_{\alpha}\rangle $
in Eq.~(\ref{001})
is defined in order to be annihilated only by the
flavor field $\nu_{\alpha L}$:
\begin{equation}
\langle 0 | \nu_{\alpha L} | \nu_\beta \rangle
\propto
\delta_{\alpha\beta}
\,.
\label{006}
\end{equation}
Let us check if this is true.
The Fourier expansion of the
left-handed components of the quantized fields of massive neutrinos is
(see, for example, \cite{Nachtmann-90})
\begin{equation}
\nu_{kL}(x)
=
\int \frac{\mathrm{d}^3\vet{p}}{(2\pi)^32E}
\sum_{h=\pm1}
\left[
a_k^{(h)}(\vet{p})
\,
u_{kL}^{(h)}(\vet{p})
\,
e^{-ip{\cdot}x}
+
{b_k^{(h)}}^{\dagger}(\vet{p})
\,
v_{kL}^{(h)}(\vet{p})
\,
e^{ip{\cdot}x}
\right]
\,,
\label{007}
\end{equation}
where $h$ is the helicity.
The neutrino and antineutrino destruction
operators
$a_k^{(h)}(\vet{p})$,
$b_k^{(h)}(\vet{p})$
satisfy the canonical anticommutation relations
(for Majorana neutrinos $a_k^{(h)}(\vet{p}) = b_k^{(h)}(\vet{p})$)
\begin{equation}
\left\{ a_k^{(h)}(\vet{p}) \, , \, {a_j^{(h')}}^{\dagger}(p') \right\}
=
\left\{ b_k^{(h)}(\vet{p}) \, , \, {b_j^{(h')}}^{\dagger}(p') \right\}
=
2E \, (2\pi)^3 \delta^3(\vet{p}-\vet{p}') \, \delta_{hh'} \, \delta_{kj}
\,.
\label{008}
\end{equation}
Hence,
\begin{equation}
\langle 0 | \nu_{kL}(0) | \nu_j(\vet{p},h) \rangle
=
u_{kL}^{(h)}(\vet{p}) \, \delta_{kj}
\,,
\label{009}
\end{equation}
and we have\footnote{
For simplicity, we consider a flavor neutrino with helicity
$h$ and definite momentum $\vet{p}$,
according with assumption \ref{A3}.
It is clear that the same conclusion is reached
considering different momenta for the massive neutrino states.
}
\begin{equation}
\langle 0 | \nu_{\alpha L}(0) | \nu_\beta(\vet{p},h) \rangle
=
\sum_{k,j}
U_{\alpha k}
U_{\beta j}^{*}
\langle 0 | \nu_{kL}(0) | \nu_j(\vet{p},h) \rangle
=
\sum_{k}
U_{\alpha k}
U_{\beta k}^{*}
u_{kL}^{(h)}(\vet{p})
\not\propto
\delta_{\beta\alpha}
\,.
\label{010}
\end{equation}
However,
since detectable neutrinos are extremely relativistic,
the contribution of neutrino masses in Eq.~(\ref{010})
can be neglected,
leading to
\begin{equation}
\langle 0 | \nu_{\alpha L}(0) | \nu_\beta \rangle
\propto
\sum_{k}
U_{\alpha k}
U_{\beta k}^{*}
=
\delta_{\beta\alpha}
\,,
\label{011}
\end{equation}
because of the unitarity of the mixing matrix.
Therefore,
the flavor states (\ref{001})
describe correctly neutrinos produced in weak interaction process
only in the extreme relativistic approximation
\cite{Giunti-Kim-Lee-Remarks-92,Bilenky-Giunti-lep-01},
which is valid in neutrino oscillation experiments.
Notice that the flavor states (\ref{001})
are not quanta of the flavor fields $\nu_\alpha$,
only appropriate linear combinations\footnote{
See Refs.~\cite{Blasone-Vitiello-99,Fujii:2001zv}
for a different point of view.
}
of the
massive neutrino states
$|\nu_{k}\rangle$,
quanta of the fields $\nu_k$.

In order to discuss the physics of neutrino oscillations
it is convenient to consider the simplest example of neutrino production:
pion decay at rest,
\begin{equation}
\pi^+ \to \mu^+ + \nu_\mu
\,.
\label{012}
\end{equation}
The energy and momentum of each massive neutrino
emitted in this process is determined by energy-momentum conservation
\cite{Winter-81}:
\begin{align}
\null & \null
p_{k}^2
=
\frac{m_{\pi}^2}{4}
\left( 1 - \frac{m_{\mu}^2}{m_{\pi}^2} \right)^2
-
\frac{ m_{k}^2 }{ 2 }
\left( 1 + \frac{m_{\mu}^2}{m_{\pi}^2} \right)
+
\frac{ m_{k}^4 }{ 4 \, m_{\pi}^2 }
\,,
\label{0131}
\\
\null & \null
E_{k}^2
=
\frac{m_{\pi}^2}{4}
\left( 1 - \frac{m_{\mu}^2}{m_{\pi}^2} \right)^2
+
\frac{ m_{k}^2 }{ 2 }
\left( 1 - \frac{m_{\mu}^2}{m_{\pi}^2} \right)
+
\frac{ m_{k}^4 }{ 4 \, m_{\pi}^2 }
\label{0132}
\end{align}
Since detectable neutrinos are extremely relativistic,
only the first order approximation in the mass contribution
is relevant:
\begin{equation}
p_k
\simeq
E
-
\xi
\,
\frac{ m_{k}^2 }{ 2 E }
\,,
\qquad
E_k
\simeq
E
+
\left( 1 - \xi \right)
\frac{ m_{k}^2 }{ 2 E }
\,,
\label{014}
\end{equation}
with
\begin{equation}
E
=
\frac{ m_{\pi} }{ 2 }
\left( 1 - \frac{ m_{\mu}^2 }{ m_{\pi}^2 } \right)
\simeq
30 \, \mathrm{MeV}
\,,
\qquad
\xi
=
\frac{1}{2}
\left( 1 + \frac{m_\mu^2}{m_\pi^2} \right)
\simeq
0.8
\,.
\label{015}
\end{equation}
Although the relations (\ref{014})
have been derived in the specific case of neutrinos
produced in pion decay at rest,
they are valid in general for any process.
Indeed,
the first order approximation in the mass contribution
must be proportional to $m_k^2$
because of the relativistic
dispersion relation (\ref{0011}).
In order to get a quantity of dimension energy,
$m_k^2$ must be divided by $E$,
that is the only available energy.
The values of $E$ and $\xi$
are determined by the production process.

From Eq.~(\ref{014})
it is clear that in general the equal momentum assumption \ref{A3}
does not correspond to reality,
unless one considers a special production process
in which $\xi=0$.
However,
as we will see in the following,
the oscillation probability turns out to be independent
of $\xi$.
Hence,
the incorrect equal momentum assumption ($\xi=0$)
leads to the correct transition probability (\ref{005}).

How to take into account the different momenta of
massive neutrinos in the derivation of the oscillation probability?
The solution to this question starts from noticing that
the oscillation probability should be Lorentz invariant
because different observers
measure the same flavor transition probability.
But the probability
(\ref{004})
is not Lorentz invariant.
In order to obtain a Lorentz invariant
oscillation probability
it is necessary to take into account not only the time evolution
of the massive neutrino states,
given in Eq.~(\ref{003}),
which depends on their energy,
but also their evolution in space,
which depends on their momentum.

The state
$ | \nu_k \rangle $
describes a massive neutrino at the production point $x=0$
at the production time $t=0$.
The state that describes the same massive neutrino at the coordinate $x$
at the time $t$
is obtained by acting on
$ | \nu_k \rangle $
with the space-time translation operator
$ e^{-iP^{\mu}x_{\mu}} $:
\begin{equation}
| \nu_k (x,t) \rangle
=
e^{-i E_k t + i p_k x}
\,
| \nu_k \rangle
\,.
\label{016}
\end{equation}
Equations~(\ref{001}) and (\ref{016})
lead straightforwardly to 
the Lorentz invariant oscillation probability
\begin{equation}
P_{\nu_{\alpha}\to\nu_{\beta}}(L,T)
=
\left|
\sum_{k}
U_{{\alpha}k}^{*}
\,
e^{ip_kL-iE_kT}
\,
U_{{\beta}k}
\right|^2
\,,
\label{017}
\end{equation}
which depends on both energy and momentum of massive neutrinos.
The probability (\ref{017})
describes oscillations in space and time.
In order to obtain the probability of oscillations in space
it is necessary to express the time $T$
in terms of the distance $L$.

In connection with the probability
(\ref{017}),
it has been claimed recently by some authors
\cite{Grossman-Lipkin-spatial-97,%
Lipkin-slit-00,%
Stodolsky-unnecessary-98}
that the equal momentum assumption \ref{A3}
should be replaced by the ``equal energy assumption''
\renewcommand{\labelenumi}{\theenumi}
\renewcommand{\theenumi}{(A\arabic{enumi}')}
\begin{enumerate}
\setcounter{enumi}{2}
\item
The massive neutrino states
$ |\nu_{k}\rangle $
have the same energy $E_k=E$,
but different momenta,
\begin{equation}
p_k
=
\sqrt{E^2 - m_k^2}
\simeq
E
-
\frac{m_k^2}{2E}
\,.
\label{0021}
\end{equation}
\end{enumerate}
This assumption
could appear to be attractive because
it leads to the vanishing of the
time dependence of the probability (\ref{017}),
which becomes a probability of oscillations in space.

From Eqs.~(\ref{014}) and (\ref{015})
one clearly see that in general the equal energy assumption
is incompatible with energy-momentum conservation
in the production process.
Therefore,
unless one considers a special production process
in which $\xi=1$,
the equal energy assumption does not correspond to reality.
However,
since the oscillation probability turns out to be independent
of $\xi$,
the incorrect equal energy assumption ($\xi=1$)
leads to the correct transition probability (\ref{005}),
as well as the incorrect equal momentum assumption ($\xi=0$).

There is another simple argument that shows that
the equal energy assumption,
as well as the equal momentum assumption,
in general do not correspond to reality:
Lorentz invariance implies that even
if different massive neutrinos have the same energy
(momentum) in one Lorentz frame,
they have different energies (momenta)
in all the other frames boosted along the
neutrino propagation path
\cite{Giunti-assumptions-01}.

Indeed,
let us assume for example that in a Lorentz frame $S$
different massive neutrinos have the same energy
$E_k = E$,
independent from the mass index $k$.
In this frame the momenta of the massive neutrinos
are given by Eq.~(\ref{0021}).

In another Lorentz frame $S'$ with velocity $v$
with respect to $S$ along the neutrino path
the energy of the $k^{\mathrm{th}}$ massive neutrino is
\begin{equation}
E'_k
=
\gamma
\left(
E + v \, p_k
\right)
\simeq
\gamma \left(1+v\right) E
-
\gamma \, v \, \frac{m_k^2}{2E}
=
E' - \frac{v}{1-v} \, \frac{m_k^2}{2E'}
\,,
\label{105}
\end{equation}
where
$\gamma = \left(1-v^2\right)^{-1/2}$
and
$
E'
=
\sqrt{\frac{1+v}{1-v}} \, E
$
is the energy of a massless neutrino in $S'$.
The difference between the energies of the
$k^{\mathrm{th}}$ and $j^{\mathrm{th}}$ massive neutrinos
in the frame $S'$
is
\begin{equation}
\Delta E'_{kj} \equiv
E'_k - E'_j
= - \frac{v}{1-v} \, \frac{\Delta{m}^2_{kj}}{2E'}
\,.
\label{107}
\end{equation}
For relativistic velocities
($v \sim 0.1 - 1$),
the energy difference is of the same order as the momentum
difference,
\begin{equation}
\Delta p'_{kj} \equiv
p'_k - p'_j
= - \frac{1}{1-v} \, \frac{\Delta{m}^2_{kj}}{2E'}
\,.
\label{108}
\end{equation}
Therefore,
it is clear that in the Lorentz frame $S'$
the energies of different massive neutrinos are different
and the equal energy assumption is untenable.

Is the transformation from a Lorentz frame $S$
to a frame $S'$ moving with relativistic velocity
with respect to $S$
important in practice?
The answer is yes.
Let us consider,
for example,
the simple case of pion decay (\ref{012}).
For the sake of illustration,
let us consider the equal energy assumption to be valid
for pion decay at rest,
even if this assumption is incompatible with
energy-momentum conservation,
as discussed above.
Then $S$ is the Lorentz frame in which the pion is at rest.

Many experiments measure the oscillations
of neutrinos produced by pion decay in flight.
These are short and long baseline accelerator experiments
and atmospheric neutrino experiments
(see \cite{BGG-review-98} for a review).
The energy of the pions goes from a few hundred MeV
(for example in the short baseline accelerator experiment
LSND \cite{LSND-flight-98})
to hundreds of GeV
(for example in the upward-going
events measured in the Super-Kamiokande atmospheric neutrino experiment
\cite{SK-upmu-99}).

It is clear that even if the equal energy assumption is valid
for pion decay at rest,
it cannot be valid even approximately
in the case of short and long baseline accelerator experiments
and atmospheric neutrino experiments.
Indeed,
considering for example
a neutrino emitted in the forward direction
by a pion decaying in flight with energy
$E_{\pi} \simeq 200 \, \mathrm{MeV}$,
the laboratory frame $S'$ is boosted with respect to the frame $S$
in which the pion is at rest by a velocity
$v \simeq 0.71$,
which gives
\begin{equation}
\frac{v}{1-v} \simeq 2.4
\,,
\qquad
\frac{1}{1-v} \simeq 3.4
\,.
\label{110}
\end{equation}
From Eqs.(\ref{107}) and (\ref{108})
one can see that the energy and momentum difference
between different massive neutrinos is of the same order of magnitude.
Obviously,
increasing the pion energy,
the energy and momentum differences increase and
tend to the same limit.

Let us emphasize that one would obtain the same result
choosing another Lorentz frame
in which the energies of different massive neutrinos
are assumed to be equal:
from Lorentz invariance
the equal energy assumption
cannot be simultaneously valid for all neutrino oscillation
experiments in which neutrinos are produced by pion decay
and it cannot be even valid in one experiment in which
the decaying pion have a spectrum of energies
(as always happens in practice).

Another obvious problem of the equal energy assumption,
as well as the equal momentum assumption,
is the arbitrariness of the choice of the Lorentz frame
in which it is valid,
which is not based on any physical argument.

Since the equal energy and equal momentum assumptions are
incompatible with Lorentz invariance and
energy-momentum conservation in the production process,
it is better to forget them
and consider the different energies and momenta of the massive neutrinos
given by energy-momentum conservation in the production process.
Using the relativistic approximations in Eq.~(\ref{014}),
the probability (\ref{017}) becomes
\begin{equation}
P_{\nu_\alpha\to\nu_\beta}(L,T)
=
\sum_k
|U_{\alpha k}|^2
|U_{\beta k}|^2
+
2
\mathrm{Re}
\sum_{k>j}
U_{\alpha k}^*
U_{\beta k}
U_{\alpha j}
U_{\beta j}^*
e^{
- i \xi \frac{\Delta{m}^2_{kj}}{2E} L
- i \left(1-\xi\right) \frac{\Delta{m}^2_{kj}}{2E} T
}
\,,
\label{018}
\end{equation}
that describes oscillations in space and time,
which depend on the characteristics of the production process
through the quantity $\xi$.

In order to obtain the probability of oscillations in space,
it is necessary to express the time $T$
in terms of the source-detector distance $L$.
This cannot be done considering massive neutrinos as plane waves,
which extend over all space at all times.
It is necessary to describe massive neutrinos as wave packets
which are localized in the production region at the production time,
propagate for a distance $L$ during the time $T$
and
are localized in the detection region at the detection time.

\begin{figure}[t!]
\begin{tabular*}{\linewidth}{@{\extracolsep{\fill}}cc}
\begin{minipage}{0.47\linewidth}
\begin{center}
\includegraphics[bb=89 384 520 773, width=0.99\linewidth]{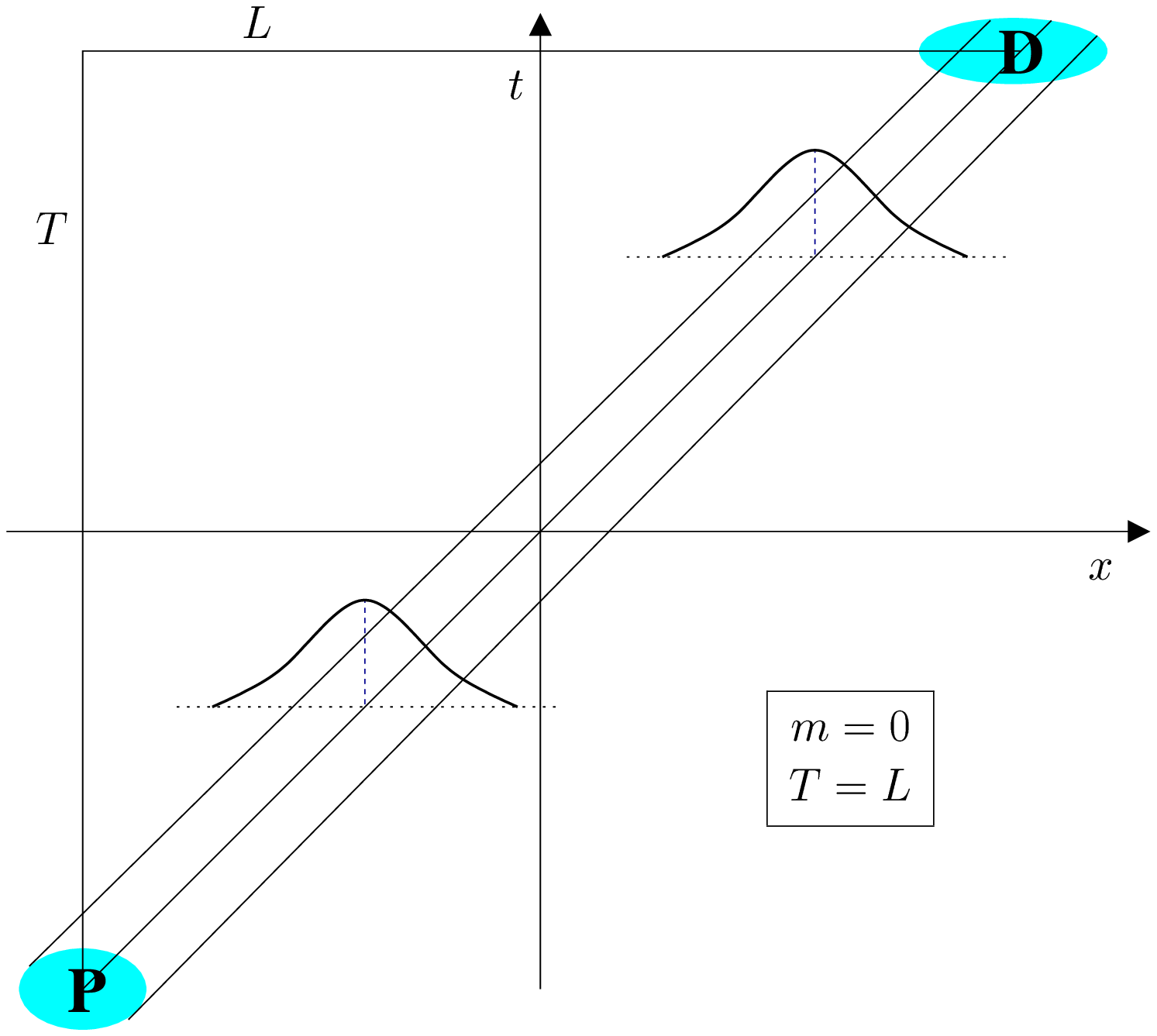}
\refstepcounter{figure}
\label{wp1}
\small
Figure \ref{wp1}
\end{center}
\end{minipage}
&
\begin{minipage}{0.47\linewidth}
\begin{center}
\includegraphics[bb=89 384 520 773, width=0.99\linewidth]{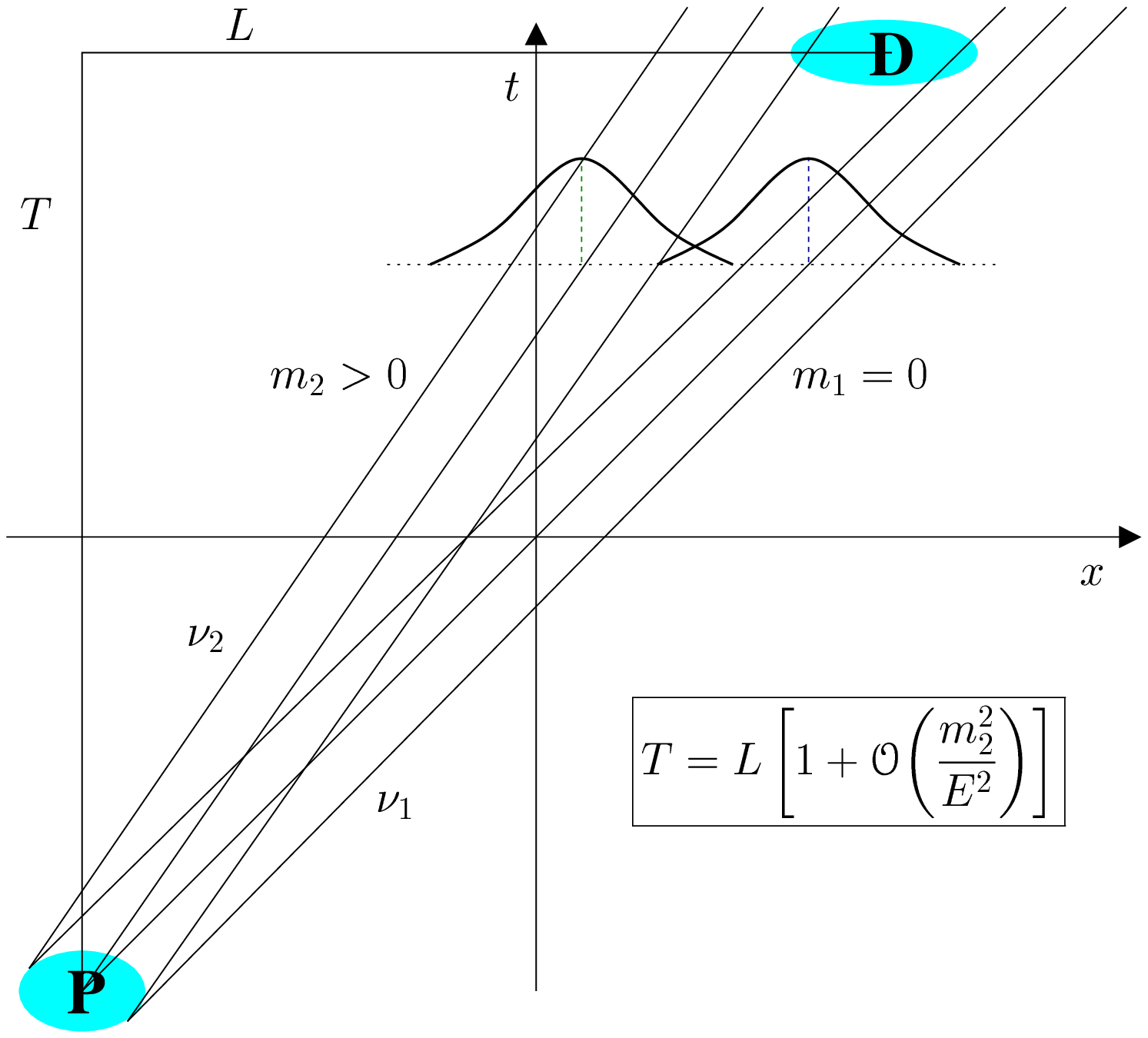}
\refstepcounter{figure}
\label{wp2}
\small
Figure \ref{wp2}
\end{center}
\end{minipage}
\end{tabular*}
\end{figure}

The production, propagation and detection of a particle wave packet
is illustrated by the space-time diagram in Fig.~\ref{wp1}.
The production and detection processes
must be localized,
but cannot be points in space-time,
because in that case the uncertainty principle would imply that
the energy and momentum have infinite uncertainty.
Therefore,
the production and detection processes occur coherently in finite
space-time regions whose sizes are connected to the
energy-momentum uncertainty by the uncertainty principle.

If the production process $P$ occurs coherently
in a finite
space-time region of dimensions
($\Delta{x}_{\mathrm{P}}$,
$\Delta{t}_{\mathrm{P}}$),
the emitted particle is described by a wave packet with
size $\sigma_x$ given by
\begin{equation}
\sigma_x^2
\sim
\Delta{x}_{\mathrm{P}}^2
+
\Delta{t}_{\mathrm{P}}^2
\,,
\label{019}
\end{equation}
as illustrated in Fig.~\ref{wp1},
in which the space-time coherence region of
the detection process has dimensions
($\Delta{x}_{\mathrm{D}}$,
$\Delta{t}_{\mathrm{D}}$).
In Fig.~\ref{wp1}
the propagating particle is assumed to be massless,
leading to the relation
$
T
=
L
$.

If only one particle propagates between the production and detection processes,
as in Fig.~\ref{wp1},
the propagating particle wave packet overlaps with both the
production and detection processes.
In this case
the wave packet description
does not lead to any interesting consequence
and can be replaced
by a plane wave treatment for all purposes except the
derivation of the relation between $T$ and $L$.
On the other hand,
in neutrino oscillation experiments
the propagating neutrino
is a coherent superposition of massive neutrinos,
each described by a wave packet,
as illustrated in Fig.~\ref{wp2}
in the simplest case of two massive neutrinos
with $m_1=0$ and $m_2>0$.
The group velocity of each neutrino wave packet
depends on its mass:
\begin{equation}
v_k
=
\frac{p_k}{E_k}
\simeq
1 - \frac{m_k^2}{2E^2}
\,.
\label{021}
\end{equation}
Different wave packets
are emitted simultaneously
by the production process,
but arrive at the detection process at different times
\begin{equation}
t_k
=
\frac{L}{v_k}
\simeq
L
\left(
1 + \frac{m_k^2}{2E^2}
\right)
\,.
\label{022}
\end{equation}
The time $T$ in the phase
\begin{equation}
\Phi_{kj}(L,T)
=
- \xi \frac{\Delta{m}^2_{kj}}{2E} L
- \left(1-\xi\right) \frac{\Delta{m}^2_{kj}}{2E} T
\label{023}
\end{equation}
of the interference term of the
$k^{\mathrm{th}}$
and
$j^{\mathrm{th}}$
massive neutrinos
in the oscillation probability (\ref{018})
must be averaged in the interval
\begin{equation}
\Delta{T}
\sim
[t_j \, , \, t_k]
=
L
+
\frac{\Sigma{m}^2_{kj}}{4E^2} \, L
\pm
\frac{\Delta{m}^2_{kj}}{4E^2} \, L
\,,
\label{024}
\end{equation}
when the wave packets of $\nu_k$ and $\nu_j$
overlap with the detection process
(for $m_k > m_j$).
In Eq.~(\ref{024}) we have defined
$\Sigma{m}^2_{kj} \equiv m_k^2 + m_j^2$.
The value of the phase
(\ref{023})
in the time interval
(\ref{024})
is given by
\begin{equation}
\Phi_{kj}(L)
\simeq
-
\frac{\Delta{m}^2_{kj} L}{2E}
-
\left(1-\xi\right)
\frac{\Delta{m}^2_{kj} L}{2E}
\left(
\frac{\Sigma{m}^2_{kj}\pm\Delta{m}^2_{kj}}{4E^2}
\right)
\,.
\label{025}
\end{equation}
The second term on the right-hand side of Eq.~(\ref{025})
is strongly suppressed with respect to the first one
because detectable neutrinos are extremely relativistic
($\Sigma{m}^2_{kj} \ll E^2$ and $\Delta{m}^2_{kj} \ll E^2$).
Moreover,
in real experiments the oscillations
due to $\Delta{m}^2_{kj}$
are observable only if
$\Phi_{kj} \sim 1$
because of the average over the neutrino energy spectrum,
the energy resolution of the detector
and the source-detector distance uncertainty.
Therefore,
the leading term
$ - \Delta{m}^2_{kj} L / 2E $
must be of order one and
the correction due to
the second term on the right-hand side of Eq.~(\ref{025})
is negligible.
In other words,
if oscillations due to
$\Delta{m}^2_{kj}$
are observable in a real experiment,
the phase $\Phi_{kj}$
is practically constant in the time interval
$ [t_j \, , \, t_k] $:
\begin{equation}
\Phi_{kj}
\simeq
-
\frac{\Delta{m}^2_{kj} L}{2E}
\,.
\label{026}
\end{equation}
As one can see from Eq.~(\ref{005}),
this is the standard value for the phase due to the
interference of the
$k^{\mathrm{th}}$
and
$j^{\mathrm{th}}$
massive neutrinos.
It is important here to notice that:
\renewcommand{\labelenumi}{\theenumi.}
\renewcommand{\theenumi}{\Alph{enumi}}
\begin{enumerate}
\item
The wave packet description of massive neutrinos
is necessary to justify the approximation $T \simeq L$
(standard assumption \ref{A4})
that leads to the standard phase (\ref{026}).
\item
The quantity $\xi$ has magically disappeared from the phase,
thanks to
the relativistic approximation and the approximation $T \simeq L$.
This is very important because it implies that
neutrino oscillations do not depend from the specific details of the
production process.
\end{enumerate}

As one can see from Fig.~\ref{wp2},
another important consequence of the wave packet description
is that the wave packets of different massive neutrinos
propagate with different velocities and tend to
separate
(it is possible to show that the spreading of
extremely relativistic wave packets is negligible).
At a distance larger than the coherence length
$L_{kj}^{\mathrm{coh}}$
the wave packets
of the
$k^{\mathrm{th}}$
and
$j^{\mathrm{th}}$
massive neutrinos
cannot both overlap with the detection process and the
corresponding interference term is suppressed
\cite{Nussinov-coherence-76}.
The coherence length
$L_{kj}^{\mathrm{coh}}$
can be estimated by equating the separation of the wave packets,
\begin{equation}
|\Delta{x}_{kj}|
=
\left|v_k-v_j\right| T
\simeq
\frac{|\Delta{m}^2_{kj}|}{2E^2} L
\,,
\label{027}
\end{equation}
to the maximal separation allowed for interference,
\begin{equation}
|\Delta{x}|_{\mathrm{max}}^2
\sim
\sigma_x^2
+
\Delta{x}_{\mathrm{D}}^2
+
\Delta{t}_{\mathrm{D}}^2
\sim
\Delta{x}_{\mathrm{P}}^2
+
\Delta{t}_{\mathrm{P}}^2
+
\Delta{x}_{\mathrm{D}}^2
+
\Delta{t}_{\mathrm{D}}^2
\,,
\label{028}
\end{equation}
leading to
\begin{equation}
L_{kj}^{\mathrm{coh}}
\sim
\frac{2 E^2}{|\Delta{m}^2_{kj}|} |\Delta{x}|_{\mathrm{max}}
\,.
\label{029}
\end{equation}

In conclusion we would like to remark that:
\renewcommand{\labelenumi}{\theenumi.}
\renewcommand{\theenumi}{\Roman{enumi}}
\begin{enumerate}
\item
The
standard expression
(\ref{005})
for the neutrino oscillation probability in vacuum is robust.
\item
The relativistic approximation is crucial
in order to obtain an oscillation probability
that does not depend on the specific details of the
production and detection processes.
\item
The oscillation probability
must be Lorentz invariant
because different observers
measure the same flavor transition probability.
\item
The equal momentum or energy assumptions
do not correspond to reality.
They are
incompatible with Lorentz invariance
and with energy-momentum conservation.
We have shown that they are not needed for the
derivation of the oscillation probability in space and time.
\item
The wave packet treatment
is necessary in order to understand
the physics of neutrino oscillations and the approximation
$T \simeq L$
that allows to obtain
the measurable oscillation probability in space
from the
Lorentz invariant
probability of oscillations in space and time.
\end{enumerate}


\end{document}